\documentclass[12pt]{article}

\usepackage{amsmath,amssymb,amsbsy,amstext, amsthm, simplewick}
\usepackage{hyperref}
\usepackage{fullpage}
\usepackage{graphicx}
\usepackage{amsmath}		
\usepackage[margin=1.0in]{geometry}
\usepackage{setspace}
\usepackage{color}
\usepackage{fancyhdr}
\usepackage{collcell}
\usepackage{datatool}
\usepackage{environ}
\usepackage{latexsym}
\usepackage{amssymb}
\usepackage{epsfig,amsmath,graphics}
\usepackage{epstopdf}
\usepackage{verbatim}
\usepackage{wasysym}
\usepackage{feynmp-auto}
\usepackage{authblk}
\usepackage{xcolor}
\usepackage{enumitem}
\usepackage[utf8]{inputenc}
\usepackage{slashed}
\usepackage{cite}
\usepackage{color}

\usepackage[framemethod=default]{mdframed}
\newmdenv[skipabove=7pt,
skipbelow=7pt,
rightline=false,
leftline=false,
topline=false,
bottomline=false,
backgroundcolor=gray!10,
linecolor=gray,
innerleftmargin=5pt,
innerrightmargin=5pt,
innertopmargin=5pt,
innerbottommargin=5pt,
leftmargin=0cm,
rightmargin=0cm,
linewidth=4pt]{eBox}

\usepackage{empheq}

    \newlength\fsep
    \newsavebox\widebox

\usepackage[skip=0pt]{caption}

\def\beq{\begin{equation}}
\def\eeq{\end{equation}}
\def\bea{\begin{eqnarray}}
\def\eea{\end{eqnarray}}

\def\0{{\vec{0}}}
\def\k{{\vec{k}}}

\begin{document}

\begin{titlepage}
\setcounter{page}{1} \baselineskip=15.5pt 
\thispagestyle{empty}
$\quad$
\vskip 50 pt

\begin{center}
{\fontsize{18}{18} \bf Neutrino Interactions in the Late Universe}
\end{center}

\vskip 20pt
\begin{center}
\noindent
{\fontsize{12}{18}\selectfont  Daniel Green$^{1}$, David E.~Kaplan$^{2}$, and Surjeet Rajendran$^{2}$}
\end{center}

\begin{center}
\vskip 4pt
\textit{ $^1${\small Department of Physics, University of California at San Diego,  La Jolla, CA 92093, USA}}
\vskip 4pt
\textit{ $^1${\small Department of Physics \& Astronomy, The Johns Hopkins University, Baltimore, MD  21218, USA}
}
\end{center}

\vspace{0.4cm}
 \begin{center}{\bf Abstract}
 \end{center}
\noindent  The cosmic neutrino background is both a dramatic prediction of the hot Big Bang and a compelling target for current and future observations.  The impact of relativistic neutrinos in the early universe has been observed at high significance in a number of cosmological probes.  In addition, the non-zero mass of neutrinos alters the growth of structure at late times, and this signature is a target for a number of upcoming surveys. These measurements are sensitive to the physics of the neutrino and could be used to probe physics beyond the standard model in the neutrino sector.  We explore an intriguing possibility where light right-handed neutrinos are coupled to all, or a fraction of, the dark matter through a mediator.  In a wide range of parameter space, this interaction only becomes important at late times and is uniquely probed by late-time cosmological observables. Due to this coupling, the dark matter and neutrinos behave as a single fluid with a non-trivial sound speed, leading to a suppression of power on small scales.  In current and near-term cosmological surveys, this signature is equivalent to an increase in the sum of the neutrino masses.  Given current limits, we show that at most 0.5\% of the dark matter could be coupled to neutrinos in this way.

\end{titlepage}

\setcounter{page}{2}

\restoregeometry

\begin{spacing}{1.4}
\newpage
\setcounter{tocdepth}{2}
\tableofcontents
\end{spacing}

\setstretch{1.1}
\newpage

\section{Introduction}
The right handed neutrino~\cite{Mohapatra:2005wg,Abazajian:2012ys,Drewes:2013gca} is a well motivated extension of the Standard Model; the non-zero mass of neutrinos implies the existence of such a particle, fundamental or composite, at some energy scale. Given that a Majorana mass for the right-handed neutrino softly breaks lepton number, it could reasonably be at or below the scale of neutrino masses, $\sim 0.1$ eV.  As a standard model singlet, such a right handed neutrino is of significant phenomenological interest since it can be a portal into a variety of dark sectors including those associated with the physics of dark matter or dark energy \cite{Bertoni:2014mva,Batell:2017cmf,Graham:2017hfr,Blennow:2019fhy,Graham:2019bfu, Berghaus:2020ekh}. The dark matter can exist at any scale while the physics directly associated with dark energy would be light, at scales comparable to the scale of neutrinos. The rich physics that could potentially be accessed through this portal  motivates the development of experimental strategies to probe light right handed neutrinos. 

The right handed neutrino interacts with the left handed neutrino by mixing with it through the Dirac neutrino mass. At high energies these interactions are suppressed, unlike typical left handed neutrino interactions that grow with energy. Existing laboratory techniques cannot overcome this suppression since there are no known methods to produce a strong source of low energy (eV scale) neutrinos. As a result, terrestrial probes of this possibility are quite limited.

Fortunately, the hot big bang provides us with such a large density of neutrinos whose impact is seen primarily through their gravitational influence.  A thermal distribution of neutrinos was created prior to big bang nucleosynthesis (BBN) and made up approximately 41 percent of the energy during the radiation era ($z>3000$).  As result, features in the power spectra generated during this era received large contributions from the gravitational effects of the cosmic neutrino background (often parameterized by $N_{\rm eff}$) and have been measured at high significance in the cosmic microwave background (CMB)~\cite{Planck:2018vyg} and relic abundances~\cite{Cyburt:2015mya}.  In addition, the gravitational influence of neutrino density fluctuations uniquely reveals that they are free-steaming at nearly the speed of light around the time of recombination~\cite{Bashinsky:2003tk,Baumann:2015rya,Baumann:2017lmt,Green:2020fjb}, an effect that has been measured in the CMB~\cite{Follin:2015hya,Baumann:2015rya,Brust:2017nmv} and large scale structure (LSS)~\cite{Baumann:2019keh}.  These physical signatures of the nature of neutrinos are responsible for the strong constraints on neutrino self-interactions~\cite{Cyr-Racine:2013jua,Park:2019ibn,Brinckmann:2020bcn} and interactions with dark matter~\cite{Wilkinson:2014ksa,Archidiacono:2014nda,Binder:2016pnr,Boehm:2017dze,DiValentino:2017oaw,Stadler:2019dii,Becker:2020hzj,Mosbech:2020ahp} prior to recombination.   

As the universe cools after the period of recombination, neutrinos become non-relativistic such that their energy density will redshift like matter. Thus, neutrinos will contribute to $\Omega_m$ as measured by the expansion history at late times.  Yet, due to their large velocities, the neutrinos do not cluster like cold dark matter or baryons, thus leading to a suppression of the matter power spectrum below the Jeans scale of the neutrinos~\cite{Lesgourgues:2006nd,Wong:2011ip,Lesgourgues:2012uu,Lattanzi:2017ubx}.  This suppression is potentially measurable in a variety of cosmological probes and provides a means to detect the sum of neutrinos masses, $\sum m_\nu$~\cite{Hu:1997mj,Kaplinghat:2003bh}. The absence of such a signal in current surveys, such as the CMB (including lensing), provides an upper limit of $\sum m_\nu < 120$ meV (95\%)~\cite{Planck:2018vyg}.  A variety of future surveys will be sensitive enough to measure a minimum sum of neutrino masses $\sum m_\nu = 58$ meV, at 2-5$\sigma$~\cite{Dvorkin:2019jgs}.  Given the long timescales relevant to cosmology, these observations also place stringent constraints on the neutrino lifetime~\cite{Chacko:2019nej,Chacko:2020hmh}.

These cosmological observations provide a unique opportunity to test  the low energy physics of neutrinos.  For $\sum m_\nu = 58$ meV, the fraction of the matter density in neutrinos is a meager $0.4\%$, and yet their effect on clustering is observable because they can move over cosmological distances at late times.  For this reason, one might expect that non-gravitational interactions could further enhance these long range effects  and thus would be strongly constrained by low-redshift cosmological observations.  Our goal is to identify generic classes of such signatures that can be accessed with current and near future cosmological data. 

In this paper, we will explore the impact of neutrino interactions on the growth of structure and the matter power spectrum.  Our focus will be on models where right-handed neutrinos interact with the dark matter or a sub-component thereof.  These interactions are dominant at low-energies, and hence low-redshifts, and may not be constrained by the primary CMB.  Nevertheless, a large exchange of momentum between the neutrinos and the dark matter will cause the system to behave as a single fluid with non-trivial sound speed in the late universe.  Such  large couplings to all the dark matter would dramatically suppress small scale power and is excluded by current data.  Interactions with sub-components of dark matter behave like an enhancement of $\sum m_\nu$ and are excluded to sub-percent levels. 

We illustrate these signatures with a concrete model. We introduce a light mediator that couples to the right handed neutrino. By itself, this mediator induces self interactions within the neutrino sector, creating scattering within the non-electromagnetic radiation species in the universe. Next, we couple this mediator to the dark matter and study the effects of the interaction between the neutrino and the dark matter. In principle, one could also couple the mediator to radiation emanating from the kinetic energy of dark energy \cite{Berghaus:2020ekh}, but these effects should be similar to those of the self-interacting neutrino effects considered here.

The rest of this paper is organized as follows. In Section~\ref{sec:model}, we discuss these models and current constraints on the parameter space and identify the experimentally accessible parts of parameter space. Following this, in Section~\ref{sec:cosmology} we describe the cosmological observables and we conclude in Section~\ref{sec:conclusions}.

\section{Model} 
\label{sec:model}

We consider a Lagrangian,
\begin{equation}\label{eq:model}
    \mathcal{L} \supset g_N \phi N N + g_{\chi} \phi \chi \chi + m^2 \phi^2 +  m_N N N + \lambda h L N + m_{\chi} \chi \chi \ ,
\end{equation}
describing the interaction of a scalar $\phi$ with a right handed neutrino $N$ of mass $m_N$ and the dark matter $\chi$ of mass $m_{\chi}$.\footnote{ In the parameter space of interest to us, it can be checked that radiative corrections to $m_N$ are smaller than $m_N$. Naively, the scalar mass $m$ would get radiative corrections. But, as a standard model singlet, there are known methods to make this coupling technically natural (for example, see \cite{Hook:2018jle}).  }. We want to know if there are interesting neutrino-neutrino or neutrino-dark matter dynamics that would be relevant for late time cosmological measurements. That is, we wish to identify the range of $g_N, g_{\chi}, m \text{ and } m_{\chi}$ so that a model with these parameters is consistent with neutrino observations in the primary CMB but lead to observable departures at later times.  As a warm up, we will first discuss the case of neutrino self interactions and then discuss the scattering between neutrinos and dark matter. The latter is observationally more interesting since it leads to qualitative impacts on the matter power spectrum. In the analysis below, for simplicity, we will work in the regime where the mass of the mediator $m$ is less than the temperature $T$ of the left handed neutrinos today. We will also choose the Dirac mass $m_N$ of the right handed neutrino to be comparable to the neutrino mass $m_{\nu}$, although all that is required for our effects is for $m_N \lessapprox m_{\nu}$. 

\subsection{Neutrino Self Interactions}\label{sec:nu_self}
The Standard Model only populates the left handed neutrino $\nu$ during the hot big bang. In our extension of the Standard Model in Equation~(\ref{eq:model}), the dominant process that disrupts standard cosmological evolution of $\nu$ is the scattering process $\nu \nu \rightarrow \phi \phi$. Specifically, efficient momentum exchange would lead to neutrino sound waves, rather than free-steaming, which is well constrained at the time of recombination~\cite{Follin:2015hya,Baumann:2015rya,Brust:2017nmv}.  This process is suppressed by the neutrino mass at high temperatures and we demand that this process is never in thermal equilibrium in the early universe when the neutrinos are relativistic (when the neutrino temperature $T_{\nu} \gg m_{\nu}$, the neutrino mass). Imposing this requirement at the surface of last scattering, we have: 

\begin{equation}
  \underbrace{ \left( \frac{g_N^4}{T_{\nu}^2} \left(\frac{m_{\nu}}{T_{\nu}}\right)^4 \right)}_{\sigma_{\nu \nu \rightarrow \phi \phi}} T_{\nu}^3 \lessapprox \underbrace{\frac{\sqrt{\rho_{\text{m}}\left(T_{LS}\right)}}{M_{pl}}}_{H_{\text{LS}}} \implies g_{N} \lessapprox \frac{T_{\nu}}{m_{\nu}} \left(\frac{\sqrt{\rho_{\text{m}}\left(T_{LS}\right)}}{M_{pl} T_{\nu}}\right)^{\frac{1}{4}}
  \label{CMBNeutrino}
\end{equation}
where $\rho_{\text{m}}\left(T_{LS}\right)$ and $T_{LS}$ are the matter density and temperature at the surface of last scattering. Taking the neutrino temperature $T_{\nu} \approx 0.2$ eV at last scattering yields a limit $g_N \lessapprox \frac{ 4 \times 10^{-8} \text{eV}}{m_{\nu}}$, similar to the bound in \cite{PhysRevD.100.103526} (see also~\cite{Cyr-Racine:2013jua,Park:2019ibn,Brinckmann:2020bcn}). When the neutrinos become non-relativistic, these interactions are no longer suppressed by the neutrino mass. The dominant process that disrupts neutrino free streaming is the scattering process $\nu \nu \rightarrow \nu \nu$ which is efficient as long as 
\begin{equation}
    \underbrace{\left( \frac{g_N^4}{m_{\nu}^2 v^4} \right)}_{\sigma_{\nu\nu \rightarrow \nu \nu}} \frac{\rho_{\nu}}{m_{\nu}} v \gtrapprox \frac{\sqrt{\rho_m}}{M_{pl}}
\label{SelfNeutrino}
\end{equation}
where $v$ is velocity of the neutrinos and $\rho_{\nu}$ the energy density in neutrinos. For non-relativistic neutrinos during matter domination $\rho_{m} \text{, } \rho_{\nu}$ redshift  $\propto (1+z)^3$, whereas $v$ redshifts as $v \sim \frac{1+ z}{1 + z_{LS}}\frac{T_{LS}}{m_{\nu}}$ until the neutrino scattering becomes important.  This process thus becomes increasingly more relevant at later values $z$ as long as $m_{\nu} v \gtrapprox m$. 

Comparing \eqref{CMBNeutrino} and \eqref{SelfNeutrino}, we see that late time cosmology at $z \sim 1$ can, in principle, probe a range of  $g_{N}$ better than the CMB by a factor: 

\begin{equation}
\frac{g^{CMB}_{N}}{g^{LSS}_{N}} \approxeq
    10 \, \left(\frac{T_{LS}}{ m_{\nu}}\right)^{\frac{3}{4}} \left(\frac{\text{eV}}{ m_{\nu}}\right)^{\frac{1}{4}}
    \label{eqn:enhancenu}
\end{equation}
taking $\frac{\rho_{\nu}}{\rho_{m}} \sim 5 \times 10^{-3}$ during matter domination and using the present day energy density in matter to be $3 \times 10^{-11} \text{ eV}^4$. Here $g^{LSS}_N$ is the smallest value of $g_N$ that could in principle be probed by late time cosmology while $g^{CMB}_N$ is the largest value of $g_N$ that is allowed by CMB constraints. 

In addition to scattering, there is also another effect that these interactions can have on neutrinos. The background density of neutrinos sources a vacuum expectation value (vev) $\langle \phi \rangle$ for the mediator and this $\langle \phi \rangle$ can change the mass $m_N$ of the right handed neutrino. This change should be smaller than $m_\nu$ in order for this analysis to be self consistent. In the early universe, when the neutrinos are relativistic, this vev is: 
\begin{equation}
\langle \phi \rangle = g_N \left(\frac{m_{\nu}}{T}\right)^2 \frac{m_{\nu}T^2}{m^2}
\end{equation}
We choose $m$ so that $g_N \langle \phi \rangle \lessapprox m_{\nu}$ yielding
\begin{equation}
    m \gtrapprox g_N m_{\nu}
\end{equation}
The only other constraint on $m$ is that it is less than the momentum transfer $m_{\nu} v
$ where these scatterings occur. Thus, by taking:
\begin{equation}
    m_{\nu} v \gtrapprox m \gtrapprox g_{N} m_{\nu}
\end{equation}
we get a consistent set of parameters. Since $g_N \ll 1$ for our entire parameter space, these constraints permit a large range of $m$ as long as we take $m_{\nu} \gtrapprox 10^{-4} $ eV. These are a consistent set of parameters where neutrino self interactions are relevant in late time cosmology while satisfying  early universe constraints.  However, as we will see in Section 3, the effects of neutrino self interactions in late time cosmology are difficult to discern observationally in the current landscape of cosmological measurements. Thus while the enhanced probe in \eqref{eqn:enhancenu} is in principle possible, it appears to be difficult to exploit it in near-term cosmic surveys.

\subsection{Neutrino Dark Matter Scattering}

Now consider the scattering of neutrinos with the field $\chi$, which makes up a fraction of the matter $f_\chi \equiv \bar \rho_\chi /\bar \rho_m$. This coupling is constrained by the CMB, like the neutrino self-interactions, by the observations that neutrinos are free-streaming during the recombination era~\cite{Follin:2015hya,Baumann:2015rya,Brust:2017nmv}. During this era, the neutrinos are both relativistic and comprise a large fraction of the energy density of the universe.  As such, we require that momentum exchange of the neutrinos is inefficient, which translates to
\begin{align}
    \underbrace{\left(\frac{g_{\chi}^2 g_{N}^2}{T_{\nu}^2} \left(\frac{m_{\nu}}{T_{\nu}}\right)^2 \right)}_{\sigma_{\nu \chi \rightarrow \nu \chi}} \frac{\rho_{\chi}\left(T_{LS}\right)}{m_{\chi}} &\lessapprox \frac{\sqrt{\rho_{\chi}\left(T_{LS}\right)/f_\chi}}{ M_{pl}} \\ 
    \implies g_{\chi} g_N &\lessapprox \frac{T_{\nu}(z_{\rm LS})^2}{m_{\nu}}  \left(\frac{m_{\chi}}{M_{pl}}\right)^{\frac{1}{2}} \left(\frac{1}{f_\chi \rho_{\chi}\left(T_{LS}\right)}\right)^{\frac{1}{4}} \ ,
\end{align}
where $T_\nu(z\approx 1100) \gg m_\nu$ at recombination follows from the current constraint $\sum m_\nu < 120$ meV.

By construction, the neutrino-dark matter cross section increases as the neutrinos cool.  After the neutrinos become non-relativistic, at $z < 100$, $\sigma_{\chi \nu \to \chi \nu} \propto v^{-4}$ where $v$ is the thermal velocity of the neutrinos.  However, by this time, the energy density of the universe is dominated by the dark matter.  As a result, the largest observational signature would arise from a large change to the dark matter momentum distribution, which occurs efficiently when
\beq\label{eqn:neutrinoDM}
\frac{\rho_\nu}{\rho_\chi} \langle \sigma_{\chi \nu \to \chi \nu} v \rangle n_\chi \gg H(T) \ .
\eeq
Using $n_\chi \propto a^{-3}$, $v \propto a^{-1}$, and $H(T) \propto a^{-3/2}$, we see the momentum exchange becomes increasing efficient as the universe expands.  As a result, we should expect novel signatures in the late-time matter distribution when\footnote{ Note that when the neutrinos are non-relativistic, but not thermalizing, the $T_\nu\propto m_\nu v$.}
\begin{equation}
   g_{\chi} g_{N} \gtrapprox m_\nu \left(\frac{T_\nu(z)^3}{m_{\nu}^3} \frac{f_\chi}{f_\nu} \right)^{1/2} \left(\frac{m_{\chi}}{M_{pl}}\right)^{\frac{1}{2}} \left(\frac{1}{f_\chi \rho_{\chi}\left(z\right)}\right)^{\frac{1}{4}} \ .
   \label{eqn:neutrinoDM2}
\end{equation}
where we have defined $f_{\nu} = \frac{\bar{\rho}_{\nu}}{\bar{\rho}_m}$. The most stringent constraints will arise from $z \approx 1$ where most current and future cosmic surveys get most of their constraining power.  Comparing the above, we see that late time cosmology can probe the combination $g_{\chi} g_{N}$ by a factor 
\begin{equation}
  \frac{ \left(g_{\chi} g_{N}\right)_{CMB}}{\left(g_{\chi} g_{N}\right)_{LSS}} \approxeq \left(\frac{f_\nu}{f_\chi} \right)^{1/2} \left(\frac{T_{\nu}(z)}{m_{\nu}
   } \right)^{1/2} \left( \frac{z_{\rm LS}+1}{z+1} \right)^{5/4} \approx \frac{40}{f_\chi^{1/2} (1+z)^{3/4}} \sqrt{\frac{\sum m_\nu}{m_\nu}}
    \label{eqn:beatcmb}
\end{equation}
better than the constraint from the CMB.  

In principle, additional constraints can be put on our model due to the fact that the dark matter sources a vev for $\phi$,
\begin{equation}
    \langle \phi \rangle = \frac{g_{\chi}\rho_{\chi}}{m_{\chi}m^2} \ ,
\end{equation}
proportional to the dark matter density.  We require this vev to not cause $\mathcal{O}\left(1\right)$ changes to the either the dark matter mass $m_{\chi}$ or the right handed neutrino mass $m_N$. Naively the vev is dominated by the earliest moments of the universe; however, the field does not track this minimum until the Hubble scale $H \lessapprox m$. Once  $H \lessapprox m$, it begins oscillating around this evolving minimum since the field will, in general, have an initial value different from this minimum. These coherent oscillations are a constituent of cold dark matter. While there is interesting dynamics that could be pursued by studying this coupled system, our goal here is to simply present a viable parameter space for late time neutrino cosmology. To that end, we make some simplifying assumptions. We assume that the dark matter component $\chi$ was relevant in the universe only from a temperature $T_i$ with an energy density $\rho_{\chi}^{i}$ that is a fraction $f_\chi$ of the total matter density $\rho_{m}$ at that time. We then demand that $g_{\chi} \langle \phi \rangle \lessapprox m_{\chi}$. For a given $g_{\chi}$ and $m$  this can be satisfied by picking a sufficiently large value of $m_{\chi}$. Similar considerations also apply to the effect of $\langle \phi \rangle$ on the right handed neutrino mass $m_N$. Note that in this analysis, the bounds on $g_N$ and $m$ that arise from self-interactions within the neutrino sector independently apply. 

It can be checked that the following example parameter space satisfies all these constraints while permitting interesting cosmological signatures. For these, we take $T_{LS} \approxeq 0.25$ eV, $T_i \approxeq \text{keV}$, 

\begin{equation*}
    g_{N} \sim  4 \times 10^{-7} 
\end{equation*}

\begin{equation*}
  10^{-7} \text{ eV}  \lessapprox  m  \lessapprox 10^{-4} \text{ eV}
\end{equation*}

\begin{equation*}
5 \times 10^4 \,  \frac{\text{eV}^2}{m} g_{\chi}\lessapprox  m_{\chi}  \lessapprox  100 \text{ TeV}
\end{equation*}

\begin{equation}
 10^{-8} \sqrt{\frac{m_{\chi}}{\text{eV} }}   \lessapprox  g_{\chi}  \lessapprox 10^{-7} \sqrt{\frac{m_{\chi}}{\text{eV} f_{\chi}}} 
\label{parameters}
\end{equation}
for neutrino masses $m_{\nu} \sim 0.1$ eV. The range of $g_{N} g_{\chi}$ shown here is narrow since we have taken $m_{\nu} \sim \mathcal{O}\left(T_{LS}\right)$. This is due to the fact that late time cosmology probes neutrino parameters better than the CMB by the ratio shown in \eqref{eqn:beatcmb}. Interestingly, unlike standard cosmological probes of neutrinos that are only sensitive to $\sum m_{\nu}$, the effects on structure formation that we discuss below are relevant even if only the lightest neutrino scatters with the dark matter. A substantially larger parameter space is thus probed if $m_{\nu} \ll T_{LS}$ for the lightest neutrino. 

In this parameter space, the dark matter self scattering is also less than the bounds from the bullet cluster~\cite{Markevitch:2003at} and thus it is possible for $\chi$ to be all of the dark matter. But, due to the $f_\chi^{1/2}$ dependence of $g_{\chi}$, these cosmological tests are sensitive to the scattering of neutrinos with even a small fraction of dark matter while still remaining perturbative. 

\section{Cosmological Measurements}
\label{sec:cosmology}

In order to understand the cosmic signatures of our low-energy neutrino interactions, we will first review the signatures of non-zero neutrino masses alone~\cite{Lesgourgues:2006nd,Wong:2011ip,Lesgourgues:2012uu,Lattanzi:2017ubx}.  We will then introduce the possibility that the neutrinos interact at low temperatures and discuss the impact on cosmological observables.  This will then allow us to translate neutrino mass constraints into constraints on neutrino interactions.

The results in this section are related to a number of previous studies of neutrino-dark matter interactions in cosmological observables~\cite{Wilkinson:2014ksa,Archidiacono:2014nda,Binder:2016pnr,Boehm:2017dze,DiValentino:2017oaw,Stadler:2019dii,Becker:2020hzj,Mosbech:2020ahp}.  In many cases, the interaction is parameterized by a constant cross-section and is thus  highly constrained by the primary CMB as well.  In contrast, the couplings to the right-handed neutrinos are enhanced at low-temperatures and are thus most important post-recombination (and thus are only weakly constrained by the primary CMB).  Nevertheless, prior work has also studied the impact on LSS~\cite{Wilkinson:2014ksa} and the similarity of the signal of interacting dark matter Standard Model to the signal of a non-zero neutrino mass~\cite{Stadler:2018dsa}.  Our goal in this section is to provide an analytic understanding of both the constraints on neutrino mass and neutrino-dark matter interactions.  This will provide a more complete understanding of how the constraints on this model depend on the details of the neutrino interactions and the particular details of cosmic surveys used to constrain them.

\subsection{Signals of Neutrino Mass}

Cosmic neutrinos begin to decouple prior to BBN, at temperatures of ${\cal O}({\rm MeV})$.  As decoupling begins when the temperature is larger than the mass of the electron, the entropy carried by electrons and positrons is converted primarily into photons and increases their temperature. The result is that the neutrino number density is diluted relative to the photons and $
T_\nu^3 \approx \frac{4}{11} T_\gamma^3$.  Much later, when the the temperature of the universe drops below the mass of the neutrinos, the energy density in neutrinos is given by $\rho_\nu = \sum m_{\nu} n_\nu$ where $n_\nu$ is the number density for a single generation of neutrinos, resulting in
\begin{align}
\Omega_\nu h^2  = 6 \times 10^{-4} \, \left( \frac{\sum m_\nu}{58 \, {\rm meV}} \right) \ .
\end{align}
For free-streaming neutrinos, the average momentum is proportional to the neutrino temperature $\langle p_\nu\rangle \approx 3 T_\nu$.  The neutrinos will be non-relativistic when $p_\nu \leq m_\nu$ and therefore, given the CMB temperature today $T_\gamma = 2.7$ K ($2.3 \times 10^{-4}$ eV), a $50$ meV neutrino would become non-relativistic at redshifts $z_{\rm NR} \approx 100$.

Once non-relativistic, the homogeneous neutrino energy density becomes a component of total (homogeneous) matter density, $\Omega_m = \Omega_{\rm cdm} +\Omega_b+ \Omega_\nu$.  During this period, matter is a dominant component of the energy density of the universe and thus $\Omega_m$ can be accurately measured from the expansion history at low redshifts via, for example, the baryon acoustic oscillations (BAO).  Of course, this measurement alone cannot distinguish these three forms of matter and thus does not constrain $\sum m_\nu$ directly.

What does distinguish neutrinos from dark matter and baryons is its large velocity.  Even though they are non-relativistic, cosmic neutrinos travel cosmological distances in the post-recombination universe.  In contrast, the dark matter and baryons are cold and their thermal velocities are largely negligible on scales larger than the largest collapsed objects (the non-linear scale)~\cite{Baumann:2010tm}.  As a consequence, neutrinos will not cluster like dark matter or baryons and thus allows for a measurement of $\Omega_\nu$ in the late universe.

We can describe the evolution of matter perturbations during this epoch by considering two fluids in the non-relativistic regime, $\rho_{\rm c} \equiv\rho_{\rm cdm} + \rho_b$ and $\rho_\nu$.  Defining $\delta_i = \delta \rho_i / \bar \rho_i$, the continuity and Euler equations become
\begin{equation}
\dot \delta_{\rm c}(\k,t)-a^{-1} k^2 u_{\rm c}=0 \qquad
\dot \delta_{\nu}(\k,t)-a^{-1} k^2 u_{\nu}=0
\end{equation}
and
\begin{align}
\dot u_{\rm c}+H  u_{\rm c} =-\frac{1}{a} \Phi \qquad 
\dot u_{\nu}+H u_{\nu} =-\frac{1}{a} \Phi-\frac{c_{\nu}^{2}}{a } \delta_{\nu} \ ,
\end{align}
respectively, where $u$ is the scalar velocity potential, $\vec v = \vec \nabla u$ and 
\beq
\nabla^2 \Phi =  4 \pi G \left(\bar \rho_c \delta_c +\bar \rho_\nu \delta_\nu \right) \ .
\eeq
These equations can be combined straightforwardly to produce the evolution equations
\beq\label{eq:dm_nu}
\begin{array}{l}
\ddot{\delta}_{\rm c}+\frac{4}{3 t} \dot{\delta}_{\rm c}=\frac{2}{3 t^{2}}\left[\beta \delta_{\nu}+(1-\beta) \delta_{\rm c}\right] \\
\ddot{\delta}_{\nu}+\frac{4}{3 t} \dot{\delta}_{\nu}=-\frac{2 \alpha}{3 t^{2}} \delta_{\nu}+\frac{2}{3 t^{2}}\left[\beta \delta_{\nu}+(1-\beta) \delta_{\rm c}\right]
\end{array}
\eeq
where
\beq
\alpha \equiv \frac{3 k^{2} c_\nu^{2} t^{2}}{2 a^{2}}=\frac{k^{2} c_{\nu}^{2}}{4 \pi G \bar{\rho}_{\rm m} a^{2}}, \quad \beta \equiv \frac{\bar{\rho}_{\nu}}{\bar{\rho}_{m}}=\frac{\Omega_{\nu}}{\Omega_{m}}  \ ,
\eeq
and we used 
\beq
a(t) \propto t^{2/3} \to H= \frac{2}{3} \frac{1}{t} \to \frac{8 \pi G}{3} \bar\rho_{\rm m} a^{-3} = \frac{4}{9} \frac{1}{t^2} \ .
\eeq
Our definition of $\alpha$ is often rewritten in terms of a free-streaming wavenumber~\cite{Bond:1983hb}, $k_{\rm fs} = \sqrt{\frac{3}{2}}\frac{aH}{c_\nu}$, such that
\beq
\alpha = \frac{3 k^{2} c_\nu^{2} t^{2}}{2 a^{2}} = \frac{2 k^2 c_\nu^2}{3 a^2 H^2} = \frac{k^2}{k_{\rm fs}^2} \ .
\eeq
As we will see, $k_{\rm fs}$ plays the role of an effective Jeans scale such that the neutrinos do not cluster when $k\gg k_{\rm fs}$.  Here $c_\nu$ is the speed of neutrino propagation, which is approximately 
\beq
c_\nu = \frac{\langle p \rangle}{m} = \frac{3 T_\nu}{m} \approx 8.7 \times 10^{-3}\, c \, \,   \frac{a_0}{a} \, \left(\frac{58 \, {\rm meV}}{\sum m_\nu}\right) \ ,
\eeq
where $a_0 =1$ is the scale factor today.  This implies that the free streaming scale is approximately,
\beq
k_{\rm fs} = 0.04 \, h \, {\rm Mpc}^{-1} \times \frac{a}{a_0} \, \left(\frac{\sum m_\nu}{58 \, {\rm meV}}\right) \label{eq:kfs_nu} \ .
\eeq
Intuitively, we expect the neutrinos will not cluster on scales $k \gg k_{\rm fs}$.

For free-streaming neutrinos, $k_{\rm fs}$ and therefore $\alpha$, is not a constant.   Clearly as we vary $k$, we are interpolating between $\alpha \ll 1$ and $\alpha \gg 1$. However, as discussed below, the correct limiting behavior in the cases $\alpha \ll 1$ and $\alpha \gg 1$ can be obtained by separately treating $\alpha$ as a constant in each regime.   When $k$ is small and $\alpha \to 0$, the equations for cold matter and neutrinos are identical and $\delta = \delta_\nu + \delta_c$ behaves just like cold matter.  On the other hand, when $k$ is large and $\alpha \gg 1$, the evolution of $\delta_\nu$ is given approximately by
\begin{align}
\ddot{\delta}_{\nu}+\frac{4}{3 t} \dot{\delta}_{\nu}\approx-\frac{2 \alpha}{3 t^{2}} \delta_{\nu} \to \delta_\nu = c t^{-\frac{1}{6}} \, \cos\left(\frac{2}{\sqrt{6}}\sqrt{\alpha} \log t \right) 
\end{align}
and therefore decays so that $\delta_\nu \to 0$.  As a result we see that the evolution of $\delta_c$ and therefore $\delta \approx \delta_c$ is also independent of $\alpha$ when $\alpha \gg 1$.  We can therefore solve these equations analytically treating $\alpha$ as a constant throughout and still reproduce the correct limiting behavior at both large and small $k$. Following~\cite{Weinberg:2008zzc}, we make the ansatz 
\beq
\delta_{\rm c} \propto t^{\gamma}, \quad \delta_{\nu}=\xi \delta_{c}
\eeq
to find that the growing modes have
\beq
\gamma =2 / 3-\frac{2 \beta \alpha}{5(1+\alpha)}, \quad \xi=\frac{1}{1+\alpha}-\frac{\beta \alpha^{2}}{(1+\alpha)^{3}}
\eeq
to linear order in $\beta \ll 1$. Combining these results, we find
\beq
\delta = (1- \beta) \delta_{\rm c} + \beta \delta_\nu = (1 +(\xi-1) \beta )\delta_{\rm c} \propto (1 +(\xi-1) \beta ) t^{\gamma} \ .
\eeq
To simplify things, we can also expand the exponential
\begin{align}
\delta(\k,\beta) 
 &\approx \delta(\k,\beta=0)  (1 +(\xi-1) \beta )  \times \left(1 -\frac{3 \beta \alpha}{5(1+\alpha)} \log \frac{a}{a_{\rm NR}} \right) \label{eq:nu_suppression} \ ,
\end{align}
 where we used $a \propto t^{2/3}$ in the matter era and $a_{\rm NR} = 1/(1+z_{\rm NR}) \approx 10^{-2}$ is the scale factor when the neutrinos become non-relativistic.
 
 To get a more accurate solution in the $\alpha = {\cal O}(1)$ regime, we can solve Equation~(\ref{eq:dm_nu}) numerically to find the result without the constant $\alpha$ approximation.  The main difference between our approximate solution and the numerical solution can be understood as follows: the scale $k_{\rm fs}$ depends on time such that it is small at earlier times and thus our suppression will appear even for $k < k_{\rm fs}(z=0)$, namely larger physical distances than the free streaming scale at redshift zero.  The physical reason is that the velocity decreases as $1/a$ which is faster than the $1/\sqrt{a}$ needed for constant $\alpha$.  As a result, the free-steaming neutrinos were travelling faster in the earlier universe and thus covered larger distances than our approximate solution assumes.

\begin{figure}[h!]
\centering
\includegraphics[width=5in]{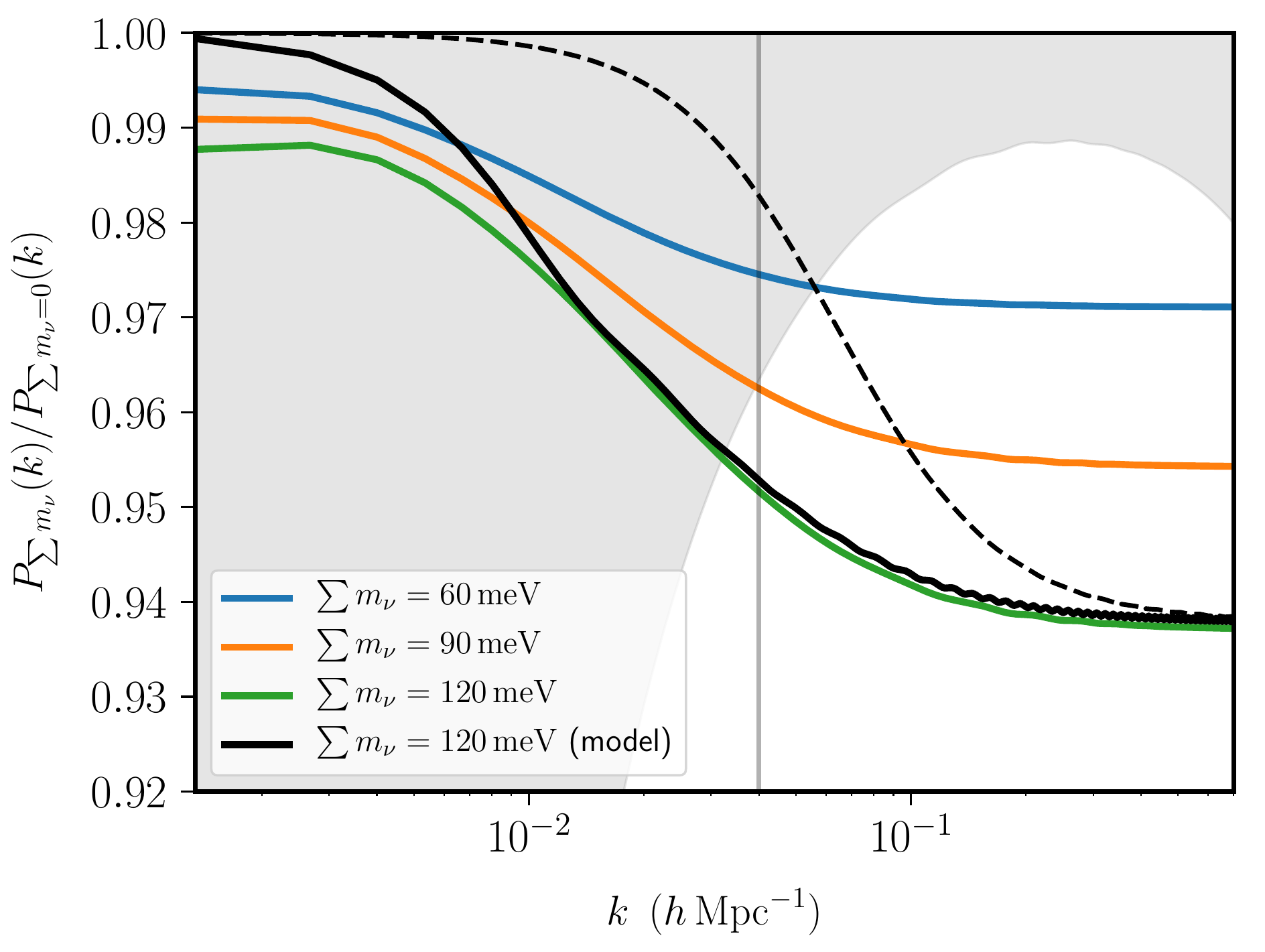}
\caption{Signal of neutrino mass in the matter power spectrum.  The colored lines indicate the results from  \texttt{CLASS}~\cite{Blas:2011rf} when varying $\sum m_\nu$ holding $H_0$, $\Omega_{\rm cdm} h^2$, $\Omega_b h^2$, and $A_s$ fixed. For comparison, the vertical line indicates the estimated free streaming scale from Equation~(\ref{eq:kfs_nu}) and the solid (dashed) black line show the result of our numeric (analytic) solution to the fluid equations for $\sum m_\nu = 120$ meV, showing excellent agreement between the fluid approximation and the Boltzmann code.  The grey band indicates the $1\sigma$ noise associated with a BOSS-like galaxy survey~\cite{Font-Ribera:2013rwa} from the combination of cosmic variance and shot noise, which dominate at low and high $k$ respectively.}
\label{fig:mnu}
\end{figure}

The influence of the non-zero neutrino mass on the matter power spectrum is shown in Figure~\ref{fig:mnu}. The solid curves show the result of calculating the matter power spectrum with CLASS~\cite{Blas:2011rf} for various choices of $\sum m_\nu$.  For comparison, we also include the numeric solution to Equation~(\ref{eq:dm_nu}) in solid black for $\sum m_\nu =120$ meV, and  our approximate analytic solution, Equation~(\ref{eq:nu_suppression}), for $\sum m_\nu =120$ meV as the dashed line.  We see three main results: first, the numeric solution  to Equation~(\ref{eq:dm_nu}) is an excellent approximation to the full results.  Second, the analytic result correctly reproduces the amplitude of the suppression and the approximate scale where the transition occurs.  Finally,  the suppression much larger than $f_\nu = \Omega_\nu / \Omega_m \approx 0.0045$ (for $\sum m_\nu = 58$ meV) since it is enhanced by the length of time that the neutrinos are non-relativistic, namely the factor of $\log a / a_{\rm NR}$ in Equation~(\ref{eq:nu_suppression}).

\subsection{Impact of Dark Matter Interactions}

Having understood the suppression of power from non-zero neutrino masses, it is easy to describe the signature of coupling the neutrinos to the dark matter and/or itself, via the portal described in Section~\ref{sec:model}.

Based on experience with dark matter-baryon interactions~\cite{Dvorkin:2013cea,Gluscevic:2017ywp,dePutter:2018xte,Gluscevic:2019yal}, we expect meaningful impacts on cosmological data only when the neutrinos scatter efficiently with  themselves or with the dark matter (or a sub component thereof).  We will focus on the interaction between the neutrinos and the dark matter sub-component in the regime where both the neutrinos and dark matter are non-relativistic.  To simplify our analysis, let us treat this as two fluids with different initial velocities / pressures that are coupled together.  As before, these interactions do nothing to the homogeneous background.  However, the fluid equations that describe the perturbations are altered,
\begin{align}
\dot \delta_\chi + a^{-1} k^2 u_\chi   &=  0 \\
\dot \delta_\nu + a^{-1} k^2 u_\nu  &= 0 \\
\dot u_\chi + H u_\chi +\frac{1}{a} \Phi &=\frac{\rho_{\nu}}{\rho_{\chi}}  R_{\chi}\left(u_{\nu}- u_\chi  \right) \\
\dot u_\nu+ H u_\nu  + c_\nu^2 k^2 \delta_\nu +\frac{1}{a} \Phi&=   R_{\chi}\left( u_\chi  - u_{\nu}\right)
\end{align}
where the momentum exchange rate is defined by
\beq
R_{\chi}=\frac{ \rho_{\chi}  } {m_{\chi}} \langle \sigma_{\chi\nu\to \chi\nu}v\rangle  \ .
\eeq
where $\langle .. \rangle$ is the average of the distribution of particles and $v$ is thermal velocity of the neutrinos.  In the tight coupling regime, $f_\nu R_\chi \gg f_\chi H$ (see Equation~\eqref{eqn:neutrinoDM}) and $u_\nu \approx u_\chi $.  We can subtract the velocity equations to eliminate $R_\chi$ to find
\beq
(1+r_\nu) (\dot u_\chi + H u_\chi ) + r_\nu c_\nu^2 k^2 \delta_\nu = -\frac{1}{a} \Phi \qquad r_\nu = \frac{\rho_\nu}{\rho_\chi}
\eeq
where we used $u_\nu \approx u_\chi $ assuming tight coupling.  Finally, when $u_\nu \approx u_\chi $ it is also easy to see that 
\beq
\frac{d}{dt}(\delta_\chi  - \delta_\nu) = 0 \ ,
\eeq
and therefore $\delta_{\chi} - \delta_\nu$ is time-independent and thus decays relative to a growing mode (i.e. one that grows in time under the force of gravity).  As a result, we can set the decaying mode to zero and substitute $\delta_{\chi} = \delta_\nu$ for the behavior of the growing mode to get a single fluid with  
\beq
c_s^2 = \frac{r_\nu}{1+r_\nu} c_\nu^2 = \frac{\rho_\nu}{\rho_\chi + \rho_\nu}c_\nu^2 \ .
\eeq
From this point forward, the impact on the growth of structure follows that same analysis as our neutrinos in Equation~(\ref{eq:nu_suppression}) with the substitutions 
\beq
c_\nu \to c_s  \qquad \text{and} \qquad \beta \to f_{\chi}  + f_\nu \, 
\eeq
where, as before, $f_{\chi} = \bar \rho_{\chi}/\bar \rho_m$ is the fraction of matter coupled to neutrinos. 

One additional impact of the scattering of neutrinos and dark matter or just neutrino self-scattering, is that the velocity of the density perturbations is no longer determined by $f(p) \approx e^{-p/T_\nu}$, as it is for free-streaming particles.  Instead, because momentum is exchanged through scattering, it is the non-relativistic kinetic energy that follows the thermal distribution. Assuming that total energy is conserved during thermalization of the neutrinos, we find 
\beq
c_\nu = \frac{\langle p \rangle}{m_\nu} = 6.7 \times 10^{-3}\, c \, \,   \frac{a_0}{a} \, \left(\frac{58 \, {\rm meV}}{\sum m_\nu}\right) \ .
\eeq
The resulting ``free-streaming" scale of the combined dark matter-neutrino fluid becomes
\beq
k_{\rm fs} \approx 0.05 \, h \, {\rm Mpc}^{-1} \times \left(\frac{f_\chi + f_\nu}{f_\nu} \right)^{1/2} \left(\frac{\sum m_\nu}{58 \, {\rm meV}}\right) \, . \label{eq:kfs_int}
\eeq
Note that even if we couple neutrinos to all of the dark matter, we have $\rho_\chi /
\rho_\nu \approx 230$ and $k_{\rm fs} \approx 0.8 \, h \, {\rm Mpc}^{-1}$. As a result, $k_{\rm fs}$ is always sufficiently small that the suppression occurs on observable scales.  Furthermore, the amplitude of the suppression is set by $\beta$ and therefore only when $f_{\chi} \approx f_\nu$ is the effect sufficiently small to not be excluded by eye.  Yet, when the fraction is small, the free streaming scale is large enough that, for the purpose of observational constraints, the suppression is equivalent to an increase in the sum of the neutrino masses.  

\begin{figure}[h!]
\centering
\includegraphics[width=3.22in]{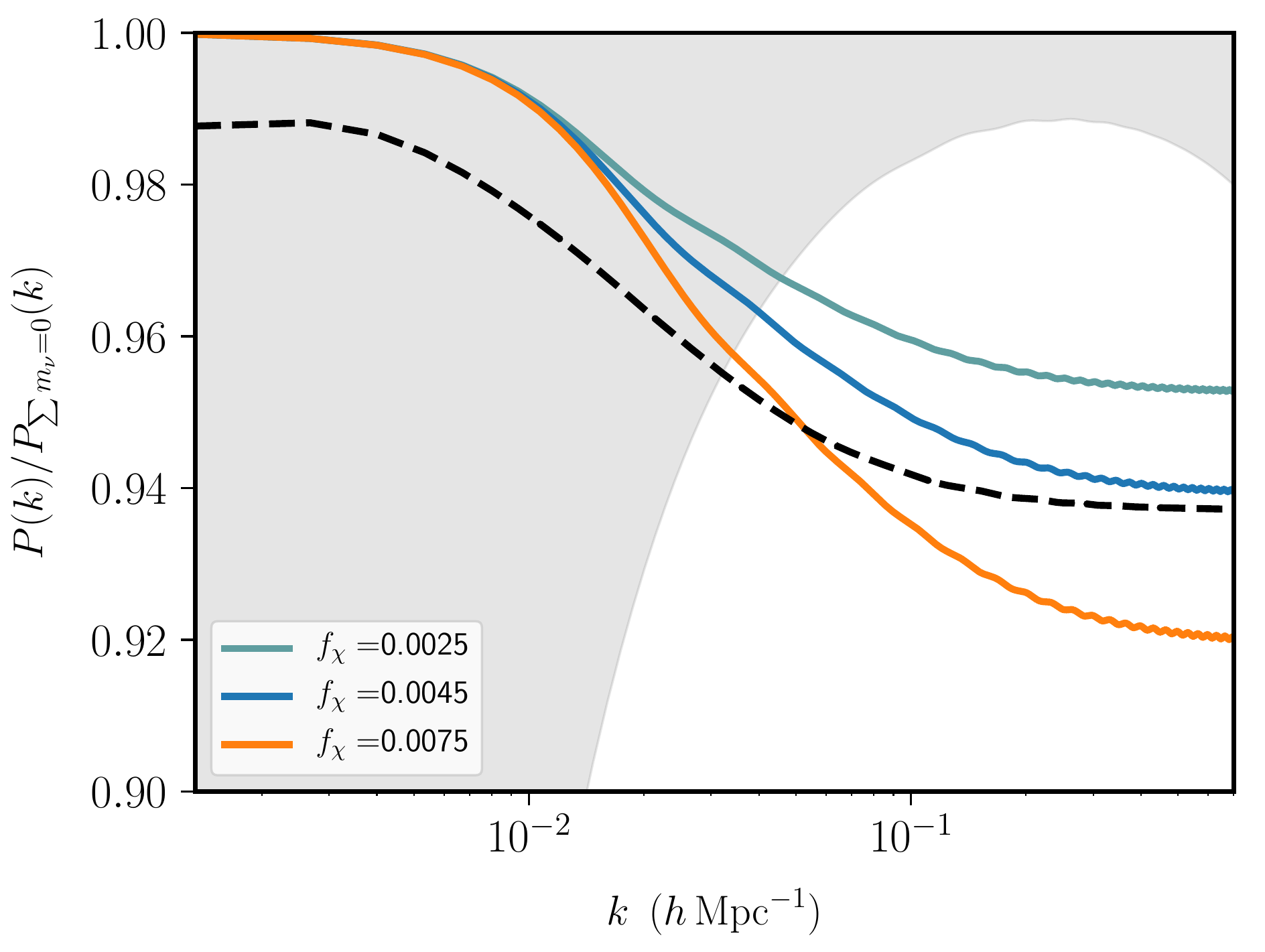}
\includegraphics[width=3.22in]{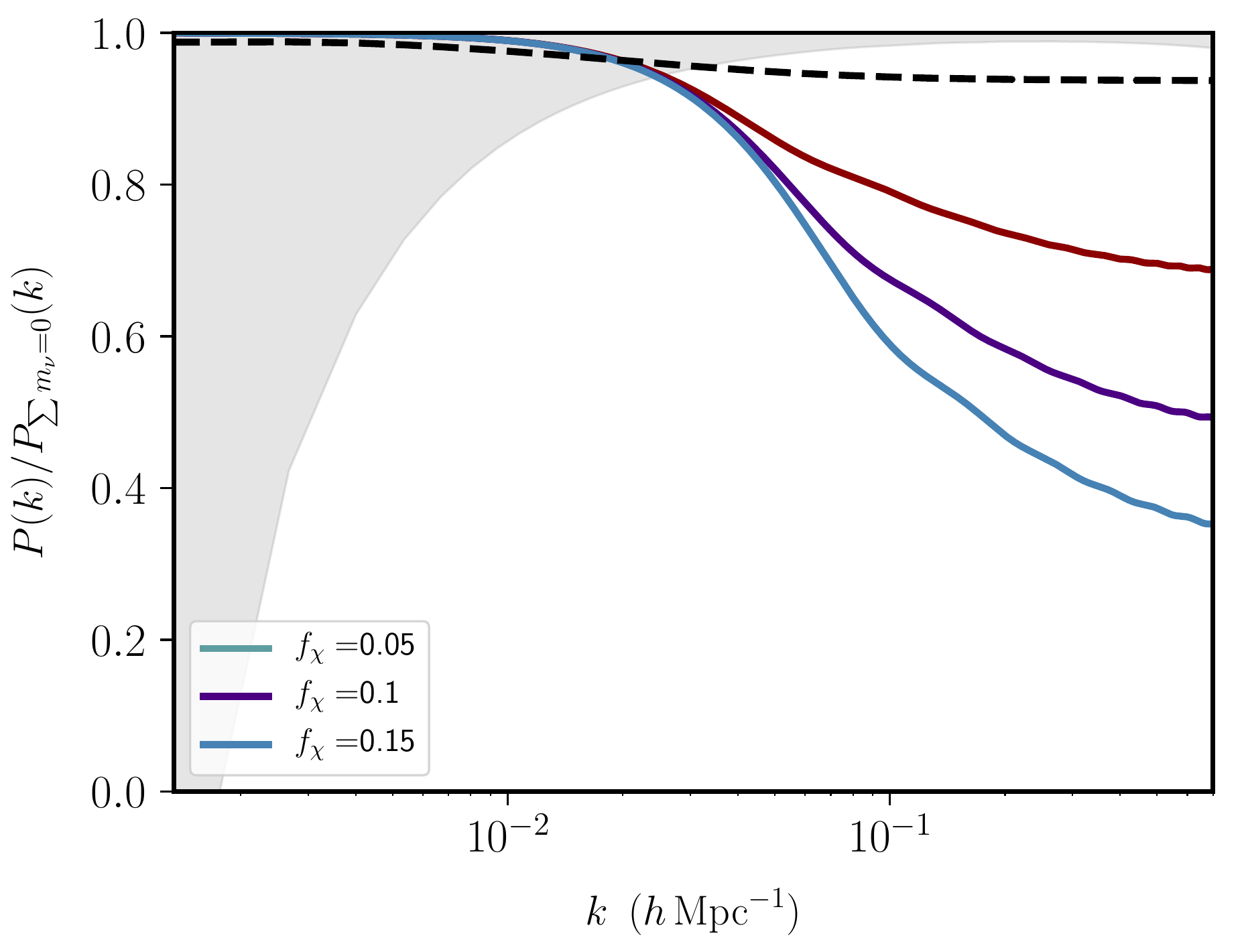}
\caption{Suppression of power due to neutrino interactions with a dark matter sub-component that makes up a fraction $f_{\chi} =\rho_{\chi} / \rho_{m}$ of the total matter density for $\sum m_\nu = 58$ meV.  The black dashed line indicates the case of $\sum m_\nu = 120$ meV (with no coupling to dark matter) for reference.  The grey bands indicate the 1$\sigma$ error of a BOSS-like galaxy survey~\cite{Font-Ribera:2013rwa} from the combination of cosmic variance and shot noise, which dominate at low and high $k$ respectively.  {\it Left:} Showing small fraction of dark matter, we conclude that $f_{\chi} <0.005$ is a conservative limit based on current constraints. {\it Right:} For larger fractions of the total energy density, we see that the suppression would be easily excluded by current observations. }
\label{fig:supp}
\end{figure}

The impact on the matter power spectrum is shown in Figure~\ref{fig:supp} when small (left) and large (right) fractions of the dark matter that are coupled to neutrinos, again using numeric solutions to Equation~(\ref{eq:dm_nu}) with the shift values of $\beta$ and $\alpha$.  The results illustrate the two most important phenomenological consequences: (1) the suppression at large $k$ is proportional to $\beta$ and (2) the free-streaming scale $k_{\rm fs}$ scales as the square-root of the fraction of the dark sectors energy that is in neutrinos.  The latter is very important, as it means that even if we coupled to all of the dark matter, $k_{\rm fs} < 1$ $h$ Mpc$^{-1}$ and therefore the suppression always occurs on observable scales.  This is shown in Figure \ref{fig:supp} by the grey region which represents the approximate noise curves of a BOSS-like~\cite{Font-Ribera:2013rwa} galaxy survey\footnote{Galaxy surveys do not constrain neutrino mass directly because a constant suppression of the small scale power is degenerate with the galaxy bias.  We could, in principle, measure the scale dependence of the suppression of the matter power spectrum at fixed redshift with such a survey but this is limited by the shown noise curves.  Lensing surveys avoid the issue of galaxy bias, but are integrated probes of the power spectrum along the line of sight and therefore combine information from a range of redshifts.}.  We see clearly that even at large $f_\chi$, due to the enhancement of the signal, that the suppression will always occur with high signal to noise.  These results are consistent with previous studies~\cite{Stadler:2018dsa} of the similarities between the signal of neutrino mass and dark matter interacting with photons. 

\subsection{Constraints on Neutrino-Dark Matter Interactions}

In order to understand the observational constraints on these dark sectors, we must first understand how $\sum m_\nu$ is constrained.  For $\sum m_\nu < 250$ meV, the neutrinos are relativistic prior to recombination and therefore their mass makes no meaningful impact on the primary CMB.  CMB lensing and/or galaxy surveys can measure that late time suppression described in Section~\ref{sec:model} through their sensitivity to the matter power spectrum.  However, as shown in Figure~\ref{fig:supp}, a common feature of these surveys is that $k_{\rm fs}$ is too small a scale to be measurable, due to the errors (cosmic variance) on the largest scales of the survey~\cite{CMB-S4:2016ple}.  As a consequence, the measure of neutrino mass for $\sum m_\nu < 250$ meV is essentially a comparison between $A_s$ measured in the primary CMB and the amplitude of the matter power spectrum on the scales observable in these surveys.  

Given that measuring neutrino mass is effectively just a comparison of two amplitudes, it is degenerate with any other parameter that alters either observed amplitude.  For the CMB, the measurement of $A_s$ is degenerate with the optical depth to reionization, $\tau$, which re-scatters the CMB photons and reduces the observed amplitude on the CMB anisotropies on scales $\ell > 30$.  It remains unclear if future CMB observations can improve on the Planck measurement of $\tau$; if not, $\sigma(\sum m_\nu)$ will be fundamentally limited to $\sigma(\sum m_\nu) > 20$ meV~\cite{CMB-S4:2016ple}.  There are several proposals that could improvement the measurement of $\tau$ by a factor of a few~\cite{Harrington:2016jrz,Smith:2016lnt,Suzuki:2018cuy,Alvarez:2019rhd}, potentially allowing $\sigma(\sum m_\nu) > 9-15$ meV~\cite{Abazajian:2019eic}.

Equally importantly, the amplitude of the matter power spectrum determined by CMB or galaxy lensing is also proportional to the total matter density, $\Omega_m h^2$.  Specifically, the suppression of power due to neutrinos can be understood as a disagreement between the values of $\Omega_m h^2$ inferred from the lensing amplitude (the amplitude of small scale fluctuations) and the expansion rate of the universe (the homogeneous matter distribution).  At present, the measurement of $\Omega_m h^2$, best determined by BOSS BAO~\cite{BOSS:2016hvq}, is also a limiting factor in the neutrino mass constraint.  Fortunately, DESI BAO~\cite{DESI:2016fyo} will substantially improve this measurement to the point that measurements are only limited by $\tau$.  

It is within this context that we can understand the current and future constraining power of cosmology on late-time dark matter neutrino interactions.  From the above discussion, we see that when the coupling of dark matter and neutrinos is sufficiently large to efficiently  exchange momentum, the observational signal is a suppression of power on small scales that is larger than would arise from the neutrinos alone.  Furthermore, while the free streaming scale is also modified by the coupling to dark matter, at most it pushes the transition into an observable range making it easier to constrain than the neutrino mass.  This is illustrated in Figure~\ref{fig:supp} by the gray band which represents the approximate noise curve of the BOSS galaxy survey~\cite{BOSS:2016hvq} based on the volume of the survey and number density of galaxies (which sets the shot noise).

\begin{figure}[h!]
\centering
\includegraphics[width=5in]{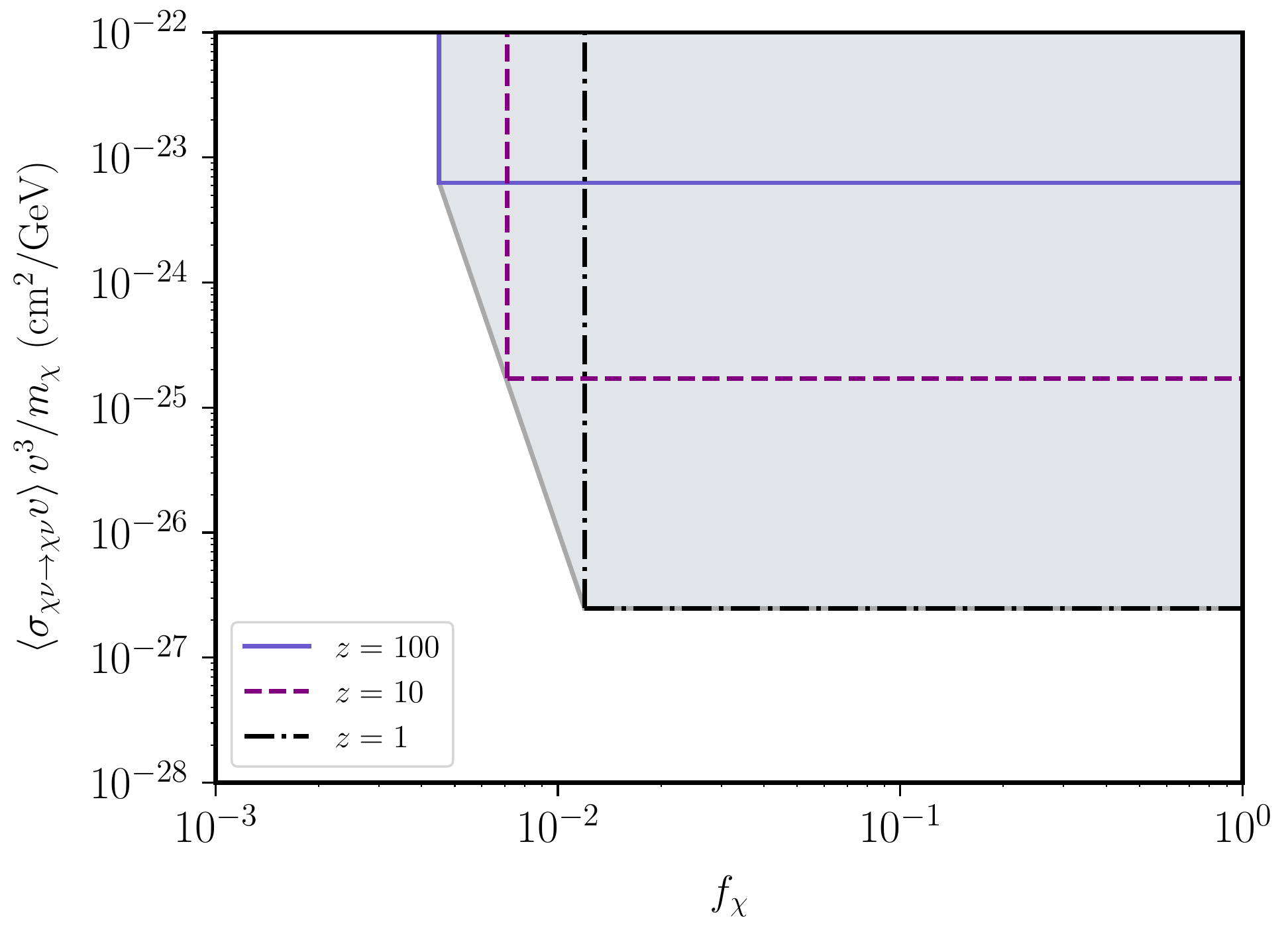}
\caption{Excluded region (95\%), in grey, in terms of the fraction of matter interacting with neutrinos, $f_\chi$, and the scattering cross section, as determined by $\langle \sigma_{\chi\nu\to \chi\nu} v \rangle v^3 \, /m_\chi$, in cm$^2$/GeV.  The grey region combines constraints on the power spectrum from efficient momentum exchanging starting at any redshift from $z=100$ to $1$.  In addition, we show the fixed-$z$ exclusions for cross-sections producing efficient momentum exchange at redshifts $z=$ 100, 10 and 1 in blue (solid), purple (dashed) and black (dashed-dot) respectively.  We see that low redshifts have more constraining power at larger $f_\chi$, as anticipated by the enhancement in Equation~(\ref{eqn:beatcmb}).  We also see that the excluded range of $f_\chi$  decreases with decreasing cross-section.  Smaller cross-sections only produce efficient scattering at lower redshifts, reducing the size of the log in Equation~(\ref{eq:nu_suppression}) and therefore the amplitude of the suppression. }
\label{fig:cross_limits}
\end{figure}

Clearly these large signatures have not been observed in current surveys, either via CMB lensing, galaxy or weak lensing surveys.  In addition, when $f_{\chi} \ll 1$, there is no meaningful observational difference between $\beta = f_\chi+f_\nu$ and $\sum m_\nu$: the transition occurs on scales where the uncertainty on measurements of the matter power spectrum are large (due to cosmic variance) and therefore only the suppression of the amplitude on small scales is observable.  Therefore, the combined constraints from current observations on our model are either that (a) the momentum exchange must be inefficient, or (b) that the fraction of matter in the coupled dark matter - neutrino fluid is smaller than the current limit on $\Omega_\nu$. These combined constraints are shown in Figure~\ref{fig:cross_limits} using Planck temperature and polarization data (TTTEEE) and lensing ($\kappa \kappa$) and BOSS BAO ($\sum m_\nu < 120$ meV at 95\%).  This figure also emphasizes that the constraining power depends on the redshift at which the neutrinos first couple efficiently to the dark matter, as described in Equation~(\ref{eqn:beatcmb}). We are more sensitive to the cross section low redshifts\footnote{The cosmological probes of neutrino mass are sensitive to the power spectrum over a range of redshifts.  For CMB lensing, the lensing kernel gets significant contributions from $z=1-5$ and therefore one might worry that coupling at $z < 5$ cannot be simply interpreted as a correction to $\sum m_\nu$.  Galaxy surveys get their information from lower redshifts and provide similar constraining power.  Moreover, this concern is largely limited to the edges of the constraints in Figure~\ref{fig:cross_limits} and do not affect the conclusions.} because of the velocity enhancement, but the effect on the power spectrum is smaller because of the shorter period over which it impacts the growth of structure (per equation~(\ref{eq:nu_suppression})). This trade-off at higher redshifts between the size of the signal and velocity enhancement manifests itself at small $f_\chi$ in this figure, where we see tighter constraints on $f_\chi$ associated with weaker constraints on the cross-section.

\begin{table}
\centering
\begin{tabular}{|l|| c | c |}
\hline
Experiment & $\sum m_\nu $ & $f_\chi$  \\
\hline
Planck~\cite{Planck:2018vyg} TTTEEE+lensing & $<0.24$ eV (95\%) & $<0.013$ (95\%)\\
+BOSS BAO~\cite{BOSS:2016hvq} & $<0.12$ eV (95\%)  & $<0.0045$ (95\%)\\
\hline
CMB-S4 TTTEEE+lensing+DESI BAO & $0.058\pm0.023$ eV (68\%) \ &$<0.0033$ (95\%)\\
+$\sigma(\tau) =0.002$ & $0.058\pm0.013$ eV (68\%) & $<0.0019$ (95\%) \\ 
+CMB-S4 Clusters & $0.058\pm0.009$ eV (68\%) &$<0.0013$ (95\%)\\
\hline
\end{tabular}
\caption{Current constraints and future projections~\cite{Abazajian:2019eic} for $\sum m_\nu$ and the derived constraint on $f_\chi$ assuming $\sum m_\nu=58$ meV is the fiducial value.  We assume the dark matter couples equally to all three neutrinos.}
\label{table:forecast}
\end{table}

Future observations can improve on these constraints by reaching smaller values of $f_\chi$ where the signal is small.  Assuming efficient momentum exchange and equal coupling to the three neutrino specifics, the upper limits on $f_\chi$ are shown for current and future CMB\footnote{Our discussion focuses on CMB lensing for comparison with the best current constraints on $\sum m_\nu$ from Planck.  Future weak lensing of galaxies with the Vera Rubin Telescope~\cite{LSSTScience:2009jmu}, for example, is expected to produce similar results~\cite{Mishra-Sharma:2018ykh}.} observations in Table~\ref{table:forecast}.  Planck data~\cite{Planck:2018vyg} with BOSS BAO~\cite{BOSS:2016hvq} gives $\sum m_\nu <$ 120 meV (95\%).  From Figure~\ref{fig:supp}, we can infer that the fraction of dark matter that can be coupled to neutrinos is $f_\chi < 0.0045$ assuming that the minimum sum of $\sum m_\nu =$ 58 meV.  If the true $\sum m_\nu$ was larger than the minimum value, the implied constraint on $f_\chi$ would be more severe.  Future CMB lensing~\cite{Abazajian:2019eic} (CMB-S4) in combination with DESI BAO~\cite{DESI:2016fyo} is expected to improve the measurement of neutrino mass to $\sigma(\sum m_\nu) = 23 $ meV; however, such a measurement would provide only a modest improvement in the constraint on $f_\chi$ down to $f_\chi < 0.033$ (95\%).  More optimistically, an independent measurement of the optical depth at the limit of cosmic variance, $\sigma(\tau) = 0.002$, and the addition of  CMB-S4 Sunyaev-Zel’dovich clusters~\cite{Madhavacheril:2017onh}, could improve the constraints down to $f_\nu < 0.0013$ (95\%).  A reasonable interpretation of these forecasts is that current data provides a remarkably powerful constraint, already limiting interactions to sub-percent level components of the dark matter, and thus improving upon this limit will take significant effort.  

Finally, let us comment on the implications for the self-interaction of the neutrinos.  As discussed in Section~\ref{sec:nu_self}, the coupling to dark matter also necessitates a neutrino self-interaction.  This interaction alone will increase $k_{\rm fs}$ slightly, as shown in Equation~(\ref{eq:kfs_int}). Unfortunately, near and medium-term surveys are unlikely to measure $k_{\rm fs}$ at a level that would distinguish this scenario for ordinary massive neutrinos~\cite{Archidiacono:2020dvx}.  Nevertheless, the free-streaming nature of neutrinos prior to recombination has been observed in the CMB~\cite{Follin:2015hya,Baumann:2015rya} and BAO~\cite{Baumann:2019keh} and thus leads to strong constraints on neutrino interactions prior to recombination~\cite{Cyr-Racine:2013jua,Park:2019ibn,Brinckmann:2020bcn}.  

\section{Conclusions}
\label{sec:conclusions}

The early universe has given us the gift of a pristine state of neutrinos with an enormous number density. This provides an ideal environment to test our understanding of neutrino physics, both in terms of the neutrino masses and new physics in the neutrino sector.  In the Standard Model, neutrinos are decoupled through all of the observable history of the universe, and thus are particularly sensitive to any new forces.  Most importantly, their large number density, comparable to the density photons, implies that physics in the neutrino sector can be deduced from their gravitational influence on cosmological observables.  This is particularly useful for right-handed neutrinos, which are otherwise very difficult to study in experimental or astrophysical settings.

In this paper, we explored the possibility that the right-handed neutrino interacts with dark matter and/or itself.  The right-handed neutrinos are only robustly populated at low temperatures and thus the interactions are only present once the neutrinos are non-relativistic, namely in the late universe.  At the same time, their velocities remain large enough to be relevant to the growth of structure.  It is this same reason that neutrino mass can be measured by late-time observables.  More dramatically, when the neutrinos are also coupled to dark matter, the late-time signal can be significantly enhanced. Like the neutrino, the physics of dark matter is also revealed through its gravitational influence.  In the presence of couplings between these sectors, the fluctuations evolve like neither dark matter nor neutrinos alone.  In near-term cosmic surveys, the largest observable impact is to enhance the suppression of power on small scales, effectively acting like a larger neutrino mass.  Current observations already tightly constrain the sum of neutrino masses and lead to sub-percent limits on the fraction of dark matter coupled to right-handed neutrinos. Interestingly, even though signal is similar to that of a larger neutrino mass, the effects discussed in this paper are sensitive to the physics of each neutrino flavor and not only the heaviest neutrinos. In particular, the effects of even a very light neutrino would be observable, significantly extending the probe of such neutrino species beyond current CMB constraints. 

The signal we describe in this paper is reminiscent of the suppression of power found in models with dark matter-baryon interactions~\cite{Dvorkin:2013cea,Gluscevic:2017ywp,dePutter:2018xte,Gluscevic:2019yal}.  In that case, the suppression arises in the early universe where the baryons are tightly coupled to photons and thus propagate at the speed of the relativistic sound waves.  This prevents the growth of structure in the dark matter, but only for modes that are subhorizon in the early universe.  As a result, the suppression occurs only on small scales.  In contrast, the suppression due to coupling to neutrinos only occurs in the late universe and thus cannot be pushed to small scales.  In principle, measurement of the matter power spectrum on even larger scales could be sensitive to the free-streaming scale itself and could better constrain neutrino self-interactions as well.  

Beyond these cosmological signatures, it is interesting to study the implications of light right handed neutrinos for the discovery of the cosmic neutrino background. The efficient mixing of the left handed neutrino with a light right handed neutrino leads to a 50 percent decrease in the rate in experiments such as PTOLEMY that aim to detect the cosmic neutrino background via the interactions of the left handed neutrino~\cite{PTOLEMY:2019hkd}. The existence of a light right handed neutrino however raises an intriguing prospect. As a standard model singlet, it is relatively straightforward to turn on new interactions between the right handed neutrino and the standard model. Due to the efficient mixing between relic left and right handed neutrinos, these interactions may potentially open a new avenue to discover the cosmic neutrino background.

\section*{Acknowledgments}

We thank Nemanja Kaloper, Marilena Loverde, Joel Meyers, and Benjamin Wallisch for discussions. DK and SR are supported in part by the NSF under grant PHY-1818899.   SR is also supported by the DoE under a QuantISED grant for MAGIS and the SQMS quantum center and the Simons Investigator award 827042. D.\,G. was supported by the US~Department of Energy under grant no.~DE-SC0019035.  We also acknowledge the use of \texttt{CLASS}~\cite{Blas:2011rf}, \texttt{IPython}~\cite{Perez:2007ipy}, and the Python packages \texttt{Matplotlib}~\cite{Hunter:2007mat} and \texttt{NumPy}/\texttt{SciPy}~\cite{Walt:2011num}.

\phantomsection
\addcontentsline{toc}{section}{References}
\small
\bibliographystyle{utphys}
\bibliography{references}
\end{document}